\documentclass[%
 reprint,
 amsmath,amssymb,
 aps,
prb,
floatfix,
]{revtex4-2}

\usepackage{amsmath,bm}
\usepackage{amsthm}
\usepackage{amssymb}
\usepackage{amsthm}
\usepackage{mathtools}
\usepackage{amsfonts}
\usepackage{graphicx}
\usepackage{xcolor}
\usepackage{enumerate}
\usepackage{parskip}
\usepackage{slashed}
\usepackage{subcaption}

\usepackage[margin=1.5cm]{geometry}   
\renewcommand{\vec}[1]{\mathbf{#1}} 
\newcommand{\ket}[1]{\left|#1\right\rangle}    

\usepackage{microtype}
\usepackage[
    colorlinks=true,
    citecolor=magenta,
    linkcolor=blue,
    urlcolor=green!50!black,
    hypertexnames=false]{hyperref}
\definecolor{urlcolor}{rgb}{0.0, 0.5, 0.0}

\begin{document} 
\title{Bond ordering in flux phases}
\author{Yifan Liu} 
\email{yliu545@ucr.edu}
\author{Vivek Aji} 
\email{vivek.aji@ucr.edu}

\affiliation{
    Department of Physics and Astronomy, University of California,Riverside, California 92521, USA 
}



\begin{abstract} 

    Contrary to canonical expectations we show that lattice translational symmetry breaking often accompanies uniformly ordered flux phases. We demonstrate this phenomena by studying a spinless-fermion model on a square latttice with nearest-neighbor repulsion.
    We find an array of flux patterns, as a function interaction strength and filling factor, that break time-reversal symmetry but may or may not preserve translational symmetry. A key finding is that the pattern of flux ordering does not uniquely determine the electronic properties. As such a class of new phases of matter are introduced that expand the candidate ground states in many body interacting systems.
\end{abstract} 

\maketitle
\section{Introduction} 

Periodic arrangement of magnetic fluxes engender new phases and phenomena in solid state systems. 
The most celebrated example is the Quantum Anomalous Hall (QAH) effect realized in the Haldane model 
\cite{Haldane}. A generalization of the model leads to the realization of the Quantum Spin Hall Effect (QSH)\cite{kanemele}. An analogous picture also explains the origin of the Quantum Hall effect in a  two 
dimensional lattice made of coupled one dimensional chains and anomalous Hall effect in Fermi liquids \cite{sun,castro,sur}. These models describe systems without any 
inter-particle interactions but other single particle effects, such as Spin Orbit Interactions (SOC), 
lead to QAH and QSH.

Flux phases also appear as possible ground states of interacting many body systems. The staggered flux phase \cite{piflux1, piflux2,piflux3}, d-density wave (DDW)\cite{ddw} and "Flux" state in the double-exchange \cite{YKM} model are examples where time reversal is broken and the unit cell is doubled. 
Translationally invariant flux phases, sometimes referred to as the Loop Current phase (LC),  
were proposed for a three band model with nearest neighbor interaction \cite{CMV,varma06, varma_rep} 
and observed in underdoped cuprates\cite{FAQ, mook,scag,kapitulnik1}. Onsite Coulomb interaction 
in a tight binding model  LaNiO$_{3}$ /LaAlO$_{3}$ hetero-structures \cite{ruegg,KYY} and nearest
 neighbor repulsion on a Honeycomb lattice also lead to flux phases displaying the QAH effect 
 \cite{raghutmi}. However the latter results have been challenged by recent density matrix renormalization studies that find a charge density wave state at half filling\cite{motruk} and a Lanczos algorithm study that finds no evidence of a topological state\cite{varney}. Recently ordered chiral orbital currents have been proposed as an explanation 
 for the colossal magnetoresistance (CMR) in Ferrimagnetic Mn$_3$Si$_2$Te$_6$\cite{yuzhang}.

The above considerations motivate a study of flux phases in general both for possible novel ground states 
and the nature of quantum phase transitions in such systems. In this letter we draw attention to an important aspect of interaction driven flux phases which has not been appreciated thus far. \textit{The spatial pattern of fluxes do not uniquely specify the electronic sector of the low energy states}. In particular lattice translational symmetry breaking is intertwined with spatially uniform intra-unit cell flux ordering. Previous studies of flux phases have not considered such states and the goal of this letter is to show that these class of states are energetically favorable among flux phases condensed by nearest neighbor repulsion. Interestingly translation breaking accompanying an LC pattern has been invoked to explain the time reversal breaking in kagome superconductors \cite{mielke}. 

To illustrate the origin of the intertwined order consider 
spinless fermions on two sites:
\begin{equation}
H = t(c_{1}^{\dagger}c_{2}+c_{2}^{\dagger}c_{1})+Vn_{1}n_{2}
\end{equation}
Mean field studies that focus on flux phases recast interaction as
\begin{equation}
Vc_{1}^{\dagger}c_{1}c_{2}^{\dagger}c_{2}=-V\phi_{12}\imath(c_{1}^{\dagger}c_{2}-c_{2}^{\dagger}c_{1})+V(n_{1}+n_{2})
\end{equation}
While the standard Hubbard-Stratonovich approach is more appropriate the important terms included in this decoupling is enough to illustrate the phenomena. 
Here $\phi_{12}$ is a real valued field. The effective Hamiltonian is
\begin{align}
    H_{\text{eff}}
    &=(t-\imath V\phi_{12})c_{1}^{\dagger}c_{2}+ (t+\imath V\phi_{12})c_{2}^{\dagger}c_{1}+V(n_{1}+n_{2})\nonumber \\
    &={t\over {\cos(a_{12})}}e^{\imath a_{12}}c_{1}^{\dagger}c_{2}+{t\over {\cos(a_{12})}}e^{-\imath a_{12}}c_{1}^{\dagger}c_{2}\nonumber \\ 
    &\qquad+V(n_{1}+n_{2})
\end{align}
where $a_{12} = \arctan\left(-{V\phi_{12}\over {t}}\right)$. Crucially there is a modification of the 
hopping matrix element that depends on the value of $\phi_{12}$ on the link. For a two site model the 
phase on the link can be eliminated by a gauge transformation and is not physically meaningful. 
However for any lattice the net flux within a plaquette, i.e. finite sum of phase field $a_{ij}$,  
cannot be eliminated. It is important to emphasize that only the phase of the link variable, 
$e^{\imath_{a_ij}}$ is affected by a gauge transformation, while the modified hopping $t/\cos(a_{ij})$ 
remains unchanged.

These considerations are analogous to mean field theories where the bond variable is a complex field $\chi_{12}$ (rather than just $\imath \phi_{12}$) whose phase was considered a redundant variable \cite{piflux1,piflux2}. However the specification of the flux pattern in each unit cell does not uniquely determine the fermionic ground state. Various orderings of $\chi_{ij}$ have been considered in the context of Heisenberg Hubbard models on a square lattice leading to a phase diagram that includes the staggered flux phase, Peierls state, and the Kite Phase. The last two are not flux phases but are bond ordered phases involving charge modulation\cite{piflux1}. 

\section{One-band model on a square lattice}
\subsection{Model} 
We consider a one-band model 
on square lattice which can spontaneously develop a loop current pattern with the
same symmetry as the $\Theta_{\text{II}}$-loop-current pattern in three-band
model\cite{allais2012loop,kivelson2012fermi,stanescu2004nonperturbative,wang2013quantum}.
The one-band model is effectively captured by the following Hamiltonian 
of spinless electrons:
\begin{align}\label{eq:Hamiltonian}
\hat H
&=-t\sum_{\langle ij\rangle}(\hat c^\dagger_{i}\hat c_j+h.c.)
-t'\sum_{\langle\langle ij\rangle\rangle}(\hat c^\dagger_{i} \hat c_j+h.c.)\nonumber\\
&\qquad+V\sum_{\langle ij\rangle}\hat n_{i}\hat n_j,
\end{align}
where $t$ and $t'$ are the nearest-neighbor (NN) and the next-nearest-neighbor
(NNN) hopping integrals, respectively. $\hat c_i$ labels the annihilation operators of electrons 
at site $\vec R_i$, and $\hat n_i$ is the corresponding density operator.
For simplicity, we only consider the nearest-neighbor interaction with strength $V$. 
The generalization to including interaction along diagonal directions is straightforward
but does not change the essential physics for our analysis.

\begin{figure}[h]
    \centering
    \includegraphics[width=0.7\linewidth]{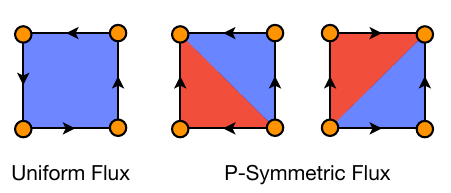}
   \caption{Possible intra-unitcell flux phases in a square lattice with next nearest neighbor hopping. 
   There is one configuration with a next flux and two with p-symmetry with next zero flux.}
   \label{fluxphase}
\end{figure}

To show how the loop-current phase can spontaneously emerge,
we perform mean-field studies of the Hamiltonian in Eq.(\Ref{eq:Hamiltonian})
and rewrite the nearest neighbor interaction by using the identity
$2\hat c_{i\sigma}^\dagger \hat c_{i\sigma}\hat c_{j\sigma'}^\dagger \hat c_{j\sigma'}
=-(i\hat c_{i\sigma}^\dagger \hat c_{j\sigma'}+h.c.)^2+\hat n_{i\sigma}+\hat n_{j\sigma'}$.
Then up to a constant proportional to the total number of electrons, the nearest neighbor interaction can be rewritten in terms of the 
bond operator $\hat J_{ij}=i(\hat c_{i}^\dagger \hat c_{j}-h.c.)$ as
\begin{equation}
    V\sum_{\langle ij\rangle}\hat n_i \hat n_j\rightarrow-\frac{1}{2}
    V\sum_{\langle ij\rangle}\hat J_{ij}^2.
\end{equation}
Our  primary interest is to establish the most energetically favorable loop current phase. Competition with other orders,
e.g. charge density wave order $\langle n_i\rangle$, bond charge density wave $\langle c^\dagger_ic_j+ c^\dagger_jc_i\rangle$,
or pair density wave $\langle c_ic_j\rangle$, and conditions for when any of the loop current phases identified here are favorable is left for future studies.
We  decouple the interaction in the ``bond current'' channel and write the mean-field Hamiltonian as 
\begin{equation}\label{eq:HStrans}
    \hat H_{\text{MF}}=\hat H_0+\sum_{\langle ij\rangle}\left[\frac{1}{2}VJ_{ij}^2
    -VJ_{ij}\hat J_{ij}\right],
\end{equation}
where $H_0$ is the free fermion Hamiltonian including the NN and NNN hoppings, and the second term in the bracket amounts to modify the nearest neighbor hopping amplitude as $t\rightarrow t+i VJ_{ij}\equiv Z_{ij}t e^{i\Phi_{ij}}$, and $\Phi_{ij}=\arg[t+i VJ_{ij}]=\tan^{-1}(VJ_{ij}/t)$, $Z_{ij}=1/\cos\Phi_{ij}$. By substitution of variable $J_{ij}\rightarrow\Phi_{ij}$, we compactify the configuration space of $J_{ij}\in\mathbb R$ to $\Phi_{ij}\in(-\pi/2,\pi/2)$. The mean field Hamiltonian in terms of $\Phi_{ij}$ fields can be written as 
\begin{align}
    \hat H_{\text{MF}}
    &=-\sum_{\langle ij\rangle}\frac{t}{\cos\Phi_{ij}}e^{i\Phi_{ij}}\hat c^\dagger_{i}\hat c_j
    - t'\sum_{\langle\langle ij\rangle\rangle}\hat c^\dagger_{i} \hat c_j +h.c.\nonumber\\
    &\qquad+\sum_{\langle ij\rangle}\frac{1}{2}VJ_{ij}^2.
\end{align}
It is worth noting that
the resulting modified nearest hopping term is reminiscent of the Peierls substitution 
for the coupling of fermions to magnetic fluxes. Consequently, the field $\Phi_{ij}$ 
resembles the gauge connection on the links, and we may interpret the gauge invariant
quantity $\phi=\sum_{(i,j)\in\triangle(\square)}\Phi_{ij}$ as the flux through the triangle(square)
plaquette. 
Since there are four links and one atom per unit cell, three independent flux configurations are possible. 
They can be classified as one with spatial s-symmetry which is a net flux through the plaquette and two 
p-symmetric configurations with net flux zero (see fig.\ref{fluxphase}). A word of caution. These labeling
of symmetries is meant to reflect the behavior under spatial rotations. 

We now discuss the possible non-equivalent ansatz that form different flux patterns. 
To constrain the number of possible ansatz, we only consider the possible translation symmetry broken
phases within the $\sqrt 2\times\sqrt 2$ unit cell and the order parameter $J_{ij}$ is assumed to be
\begin{equation}
    J_{ij}=\sum_{\nu=1}^4 J_\nu\delta_{\vec e_\nu,\vec e_{ij}}
\end{equation}
where $i\in A$ sublattice, $j\in B$ sublattice, $\vec e_1=-\vec e_3=\vec e_x$ and $\vec e_2=-\vec e_4=\vec e_y$.
$\vec e_{ij}$ is the unit vector connecting site $i$ and $j$. The mean field Hamiltonian can then be written in terms of  2-component spinor $C_\vec k=(c_{\vec kA},c_{\vec kB})^T$
in the Fourier space,
\begin{eqnarray}
    \frac{\hat H_{\text{MF}}}{N}
    &=&\frac{V}{4}\sum_{\nu=1}^4 J_\nu^2
    +\frac{1}{N}\sum_{\vec k,s=\pm}'
    \hat C_\vec k^\dagger h_\vec k \hat C_\vec k,\\ 
    h_\vec k
    &=&
    \begin{pmatrix}
        -4t'c_xc_y & -\sum_{\nu=1}^4t_\nu e^{i\vec k\cdot\vec e_\nu} \\
        -\sum_{\nu=1}^4t_\mu e^{i\vec k\cdot\vec e_\nu} & -4t'c_xc_y 
    \end{pmatrix}
\end{eqnarray}
where $c_x\equiv\cos k_x$, $c_y\equiv\cos k_y$. Here $t_\nu=t+iVJ_\nu$ and $\sum_{\vec k}'$ sums over the states in the first Brillouin zone of size $\sqrt{2} \pi\times\sqrt{2} \pi$ as the unit cell is doubled. Diagonalizing $\hat H_{\text{MF}}$ by a unitary transformation, 
\begin{align}
    \begin{pmatrix}
        \hat c_{\vec kA} \\
        \hat c_{\vec kB}
    \end{pmatrix}
    =\frac{1}{\sqrt{2}}\begin{pmatrix}
        e^{i\theta_{\vec k}/2} & -e^{i\theta_{\vec k}/2} \\
        e^{-i\theta_{\vec k}/2} & e^{-i\theta_{\vec k}/2}
        \end{pmatrix}
    \begin{pmatrix}
        \hat \alpha_{\vec k+} \\
        \hat \alpha_{\vec k-}
    \end{pmatrix}
\end{align}
with $\theta_\vec k=\arg\left(\sum_{\nu=1}^4t_\nu e^{i\vec k\cdot\vec e_\nu}\right)$, we obtain
\begin{align}
    \frac{\hat H_{\text{MF}}}{N}=\frac{V}{4}\sum_{\nu=1}^4 J_\nu^2
    +\frac{1}{N}\sum_{\vec k,s=\pm}'\epsilon_{\vec ks}
    \hat\alpha^\dagger_{\vec ks}\hat\alpha_{\vec ks},
\end{align}
where  $s$ is a band index and the dispersions are given by 
\begin{align}
    \epsilon_{\vec ks}=-4t'\cos k_x\cos k_y+s\left|\sum_{\nu=1}^4t_\nu e^{i\vec k\cdot\vec e_\nu}\right|.
\end{align} 
For a given temperature and filling factor $\nu=\frac{1}{N}\sum_ic^\dagger_i c_i$, the phase will be favored with the lowest free energy in the canonical ensemble
\begin{align}
    \frac{\mathcal E_{\text{MF}}}{N}=\frac{V}{4}\sum_{\mu=1}^4 J_\mu^2
    +\frac{1}{N}\sum_{\vec k,s=\pm}'\epsilon_{\vec ks}
    n_F(\epsilon_{\vec ks}),
\end{align}
The chemical potential determined through $\frac{1}{2}\sum_{\vec ks}\Theta(\mu-\epsilon_{\vec ks})=N\nu$,
where $\Theta$ is the step function and the sum is performed over the first Brillouin zone of the underlying square lattice of size $2\pi\times 2\pi$.

\subsection{Mean field phase diagram}
We performed numerical calculations on a square lattice of size $N =100\times100$ at zero temperature.  Notice that the total energy is symmetric under $G=\{D_4,D_4T\}$ operations of $(J_1,J_2,J_3,J_4)$,  i.e. $\mathcal E_{\text{MF}}[J_1,J_2,J_3,J_4]=\mathcal E_{\text{MF}}[(J_1,J_2,J_3,J_4)^g]\ (g\in G)$, where $D_4$ is the dihedral group of a square, $T$ is the time-reversal operation that flips the sign of all $J$'s, and $(J_1,J_2,J_3,J_4)^g$ is the action of a group element $g$ on the tuple $(J_1,J_2,J_3,J_4)$. To reduce the search space, we impose symmetry breaking constraints that retain only one representative (flux pattern) in each symmetry class. This approach leverages lexicographic ordering constraints (LOC) \cite{gent2006symmetry,goldsztejn2015variable} as a method for symmetry reduction.
\subsubsection{Lexicographic order}
For two vectors $x,y$ in $\mathbb R^n$ the lexicographic order is defined inductively as
\begin{align}
    (x_1,...,x_n)&\preceq_{lex}(y_1,...,y_n)\equiv (x_1< y_1)\vee \nonumber\\
    &\left((x_1= y_1)\wedge (x_2,...,x_n)\preceq_{lex}(y_2,...,y_n)\right).
\end{align}
Given a symmetry $g\in G$ of the parameter space, we define the corresponding symmetry-breaking constraint as
$\textup{LEX}_g(x)=x\preceq_{lex} x^g.$ For example, let $r$ be an rotation by $\pi/2$,
then $\textup{LEX}_r(J_1,J_2,J_3,J_4)=(J_1,J_2,J_3,J_4)\preceq_{lex}(J_4,J_1,J_2,J_3).$
The complete symmetry-breaking constraint then is the conjunction of all $\textup{LEX}_g(x)$, i.e.
$\textup{LEX}_G(x)=\bigwedge_{g\in G}\textup{LEX}_g(x).$   For instance, with $D_4$ symmetry,
\begin{align}
    &\textup{LEX}_{D_4}(J_1,J_2,J_3,J_4)=
    (J_1 = J_2 \le J_3\le J_4 )\vee\nonumber\\
    &\quad( J_1\le J_3<J_2\le J_4) \vee (J_1 < J_2 \le J_3\wedge J_2\le J_4).
\end{align} 
The reduced search space then is $S_G=\{x\in \mathbb R^n|\textup{LEX}_G(x)\}$. For finite groups with large order, managing multiple complex orders makes them impractical for numerical constraint satisfaction problems. However, this complexity associated the boundaries of the reduced search space can be simplified, given that the boundaries are normally negligible in numerical domains. To do this, we follow Ref.\cite{goldsztejn2015variable} and consider the relaxed LEX constraint defined as $\textup{RLEX}_g(x)\equiv (x)_{n_g}\le (x^g)_{n_g}.$ Here 
$(x)_{n_g}$ is the $n_g$-th component of $x$ and is the smallest number that satisfies $(x)_{n_g}\neq(x^g)_{n_g}$. For a rotation by $\pi/2$, $\textup{RLEX}_r(J_1,J_2,J_3,J_4)=J_1\le J_4;$ and for a rotation by $\pi$ around $x$-axis $\sigma_x$, $\textup{RLEX}_{\sigma_x}(J_1,J_2,J_3,J_4)=J_4\le J_2.$ It turns out that 
the corresponding search space generated by $\textup{RLEX}$ order is the closure of that generated by $\textup{LEX}$ order \cite{goldsztejn2015variable}. Therefore, we have the following relaxed constraints:
\begin{widetext}
\begin{align}\label{eq:rlex}
    &\textup{RLEX}_{G}(J_1,J_2,J_3,J_4)=
    (J_1 \le J_2 \le J_4 )\wedge (J_1\le J_3)\wedge (J_1 \le 0) \wedge (J_2 \le -J_1) \wedge (J_3 \le -J_1) \wedge (J_4 \le -J_1).
\end{align}
\end{widetext}

\subsubsection{Phase diagram}

\begin{figure}[ht]
    \centering
\begin{subfigure}{0.8\columnwidth}
  \centering
  \includegraphics[width=1\linewidth]{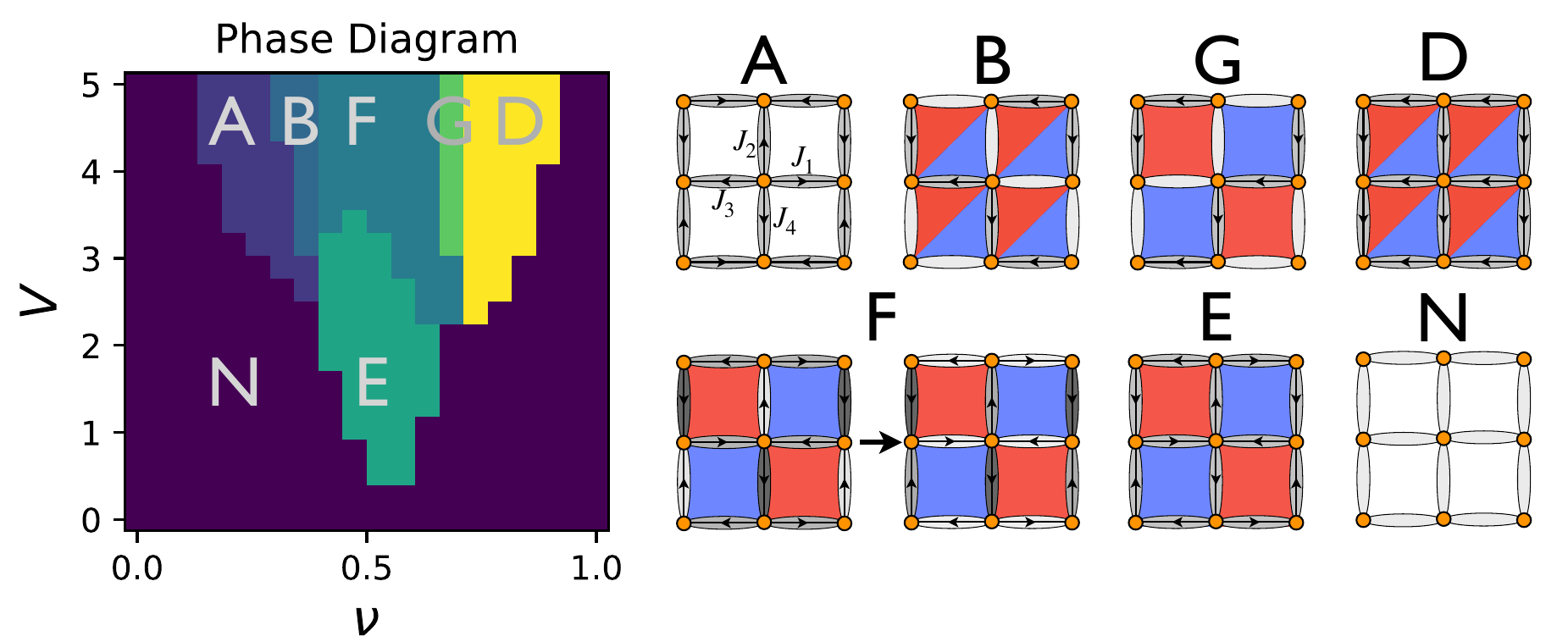}  
  \caption{Phase diagram}
  \label{fig:Phase_diagram1}
\end{subfigure}
\centering
\begin{subfigure}{0.8\columnwidth}
  \centering
  \includegraphics[width=1\textwidth]{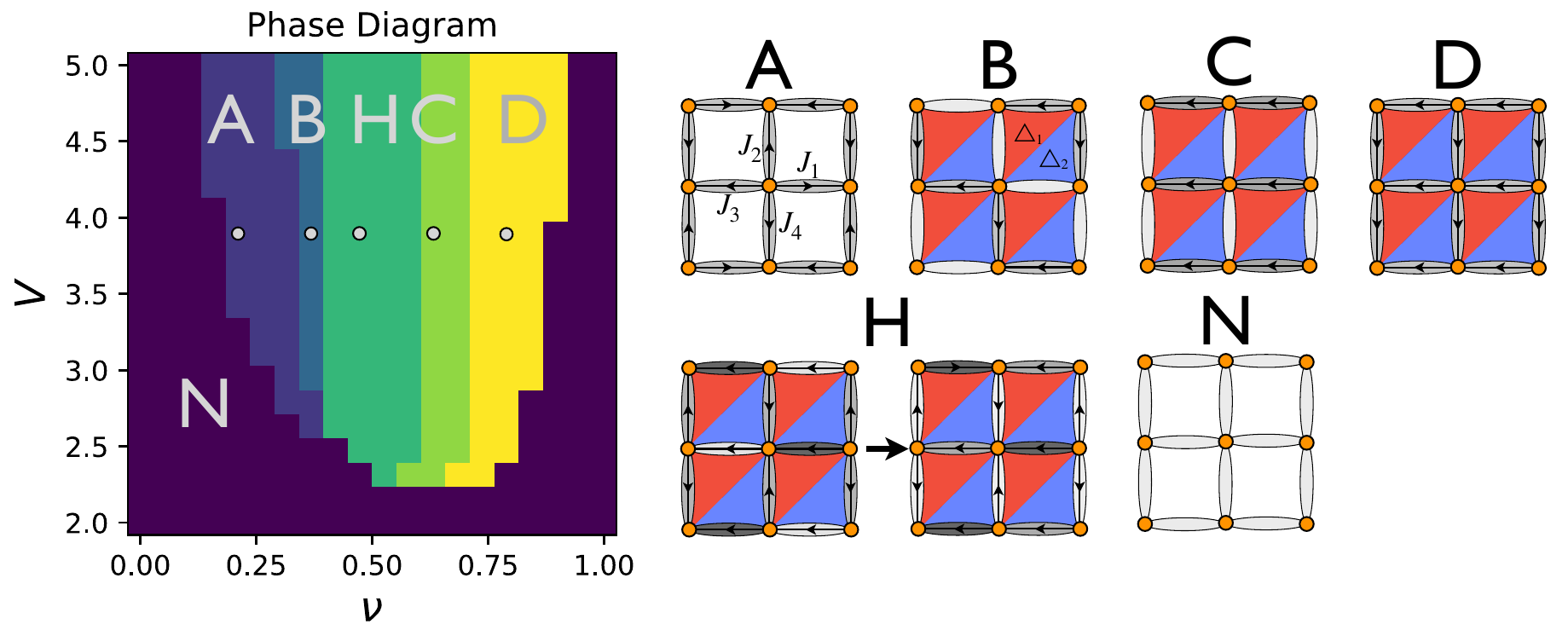}
    \caption{
    Phase diagram projected onto the zero flux subspace.
    }
    \label{fig:Phase_diagram2}
\end{subfigure}
\caption{Mean-field phase diagram in the ($\nu,V$) plane at fixed $t=1$ eV, $t'=0.2$ eV. The arrow denotes the direction of $J_{ij}$, and the gray scale represents the magnitude of $J_{ij}$.}
\label{fig:Phase_diagram}
\end{figure}

In Fig. \ref{fig:Phase_diagram}, the zero-temperature mean field phase diagram is presented, obtained by minimizing the total energy with respect to the parameters $(J_1, J_2, J_3, J_4)$ across various doping levels and interaction strengths. This optimization utilizes the Simplicial Homology Global Optimization (SHGO) algorithm \cite{endres2018simplicial}, combined with the constraints in 
\ref{eq:rlex}. We set $t=1$ eV, $t'=0.2$ eV and the cutoff strength of $V$ to be around 5 eV, above which $V$ exceeds the the band-width. We examine two scenarios: one with a finite net flux within the square plaquette, and the other without.

\begin{figure}[ht]
  \centering
  \includegraphics[width=0.8\columnwidth]{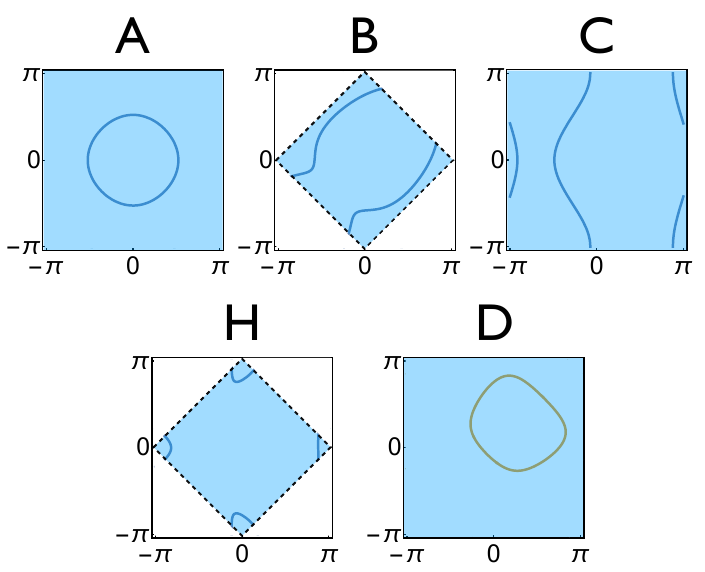}
    \caption{ 
    Typical Fermi surfaces at gray dots in the zero-flux phase diagram. The shaded area denotes the Brillouin zone.
    }
 \label{fig:Fermi_surfaces}
\end{figure}

\textbf{Phase diagram 1:}
Our result shows a rich phase diagram over wide range of the ($V,\nu$) plane. The phases
can be divided to three classes. 
(1) Fermi liquid states without flux which include phases A and N: Phase A has all $J_{\text{MF}}$'s being the same on the links but it still corresponds to zero flux state.
This is because the relative phase that appears in the effective hoppings $t+iJ_{\text{MF}}=t/\cos(\Phi) e^{i\Phi}$ can be removed by a gauge transformation $c_{iA(B)}\rightarrow e^{i\theta_{A(B)} }c_{iA(B)}$  with $\theta_{B}-\theta_{A}=\Phi$. Moreover, phase A does not break any lattice symmetry  as the magnitude of $J_{\text{MF}}$ (and thus the effect hopping) is the same everywhere.
(2) $\Theta_{\text{II}}$ (see \cite{varma06}) like phases which include phases B, G and D:
These three phases have the same symmetries as the $\Theta_{\text{II}}$ loop current order in the flux sector. The presence of flux within triangle plaquettes explicitly breaks time reversal symmetry and $C_4$ symmetry, while the lattice translational symmetry is respected by the flux. Although phases B and G have translational symmetry in the flux sector, they break the translational symmetry in the distribution of $J_{ij}$ on the links. This has consequences for the electronic sector and the symmetry of of the lattice. Phase D corresponds to the canonical $\Theta_{\text{II}}$ (see \cite{varma06}) preserving lattice transaltional symmetry in all sectors.
(3) Staggered flux phases that include phases E, F and G: Phase E has the same magnitude of $J_{ij}$ on the links, and it explicitly breaks time reversal symmetry and also breaks $C_4$ down to $C_2$. Phase F is a collection of phases with one $J$ in $(J_1, J_2, J_3, J_4)$ being the largest. Phase G is characterized by two non-vanishing $J_{ij}$'s on the links in the enlarged unit cell which resembles the stripe phase.

\textbf{Phase diagram 2:}
We now focus on the subspace of flux phases with vanishing net flux per unit cell. This 
represents the cases where additional terms in the Hamiltonian are present, such as 
$H'\sim U(\sum_{(i,j)\in\square}\Phi_{ij})^2$, which restrict the low energy sector to the 
p-symmetric states. Our numerical result shows that the previous staggered flux phases 
now are replaced by phase C and H near the half filling $\nu=0.5$. Phase C respects  
lattice translational symmetry for the flux but has stripe-like bond order $J_{\text{MF}}$. Phase H represents a collection of phases where one bond order $J$ in $(J_1, J_2, J_3, J_4)$ is the largest.

\subsubsection{Fermi surfaces} 
In this section, we discuss the evolution of Fermi surfaces within the zero-flux phase diagram at a fixed interaction strength ($V\approx 4$ eV) as the filling factor is varied. Our results are shown in Fig. \ref{fig:Fermi_surfaces} for the ground state in a single domain. Phase A does not break any symmetry but has renormalized hopping. Hence the symmetry of the Fermi surface is unchanged. Phases B and C both break translational and rotational symmetries. They have stripe-like ordering resulting in open Fermi surfaces. Phase H leads to electron pockets at $(\pi,0)$ and $(0,\pi)$. Phase D is translationally invariant but has broken rotational symmetry. These results underscore the central point of this paper which is that the symmetry of the flux sector within unit cells do not uniquely determine the ground state of the system.

\subsection{Discussion and Summary} 

In this work, we have studied a spinless-fermion model on a square lattice, in which the nearest-neighbor interaction can be written in terms of bond-current operators, giving rise to various flux phases that breaks  time-reversal. Contrary to previous studies on flux phases, we have shown in this simple model that flux pattern cannot uniquely determine the fermionic ground state. This main conclusion calls for further studies of the validity of the arbitrary assignments of gauge fields on the links for flux phases driven by interaction. An open question to be studied in the future is the competitiveness of this class of states with other forms of symmetry breaking in the charge and spin sector. Future studies will report on lattice models which include nearest neighbor repulsion using methods that better capture fluctuations beyond mean field.

\bibliography{main.bib}

\end{document}